# Radiation pressure and the distribution of electromagnetic force in dielectric media

Masud Mansuripur

College of Optical Sciences, The University of Arizona, Tucson, Arizona 85721

**Abstract**: A detailed distribution of the force of electromagnetic radiation in and around dielectric media can be obtained by a direct application of the Lorentz law of force in conjunction with Maxwell's equations. We develop a theory of the force exerted by a focused light beam on the free surface as well as within the volume of a transparent dielectric medium. Although the medium can be either solid or liquid, here we emphasize the application of the formulas to liquids since, in principle at least, surface deformations and liquid motions are measurable. Our theory predicts that, upon entering the liquid from the free space, the beam of light exerts an outward vertical force on the entrance surface that tends to produce a localized bulge. This surface force, however, is much weaker than that predicted by prevailing theories and, in contrast to current beliefs, is found to depend on the polarization state of the incident beam. Within the volume of the liquid we predict that the forces of radiation tend to create four counter-rotating vortices at the four corners of the focused spot; the sense of rotation within these vortices depends on the direction of the incident polarization. These striking departures from conventional wisdom with regard to the force of radiation arise from a revision in the form of the Lorentz law as applied to the bound charges/currents within a dielectric medium.

**Keywords**: Radiation pressure; Optical trapping; Optical binding, Electromagnetic theory.

**1. Introduction**. The 1973 experiments of Ashkin and Dziedzic on purified water were claimed to have demonstrated the formation of a bulge on the water surface at the point of entry of a focused laser beam [1]. The mechanism of the force responsible for the bulge is no longer believed to be the increase of the photon momentum from its free-space value of $\hbar\omega/c$ to the Minkowski value of $n\hbar\omega/c$ for light of frequency $\omega$ in a medium of refractive index $n$, as was initially suggested in [1]. In recent years Loudon, who has developed a comprehensive quantum theory of radiation pressure [2, 3], has attempted to explain the observations of Ashkin and Dziedzic by starting from the Lorentz law of force [4]. In Loudon's treatment, the $E$-field contribution to the Lorentz force is expressed as $\boldsymbol{F}_e = (\boldsymbol{P}\cdot\nabla)\boldsymbol{E}$, where $\boldsymbol{P} = \varepsilon_\mathrm{o}(n^2 - 1)\boldsymbol{E}$ is the induced polarization density by the local electric-field $\boldsymbol{E}$ within a dielectric medium of index $n$. Together with the magnetic-field contribution, $\boldsymbol{F}_m = (\partial\boldsymbol{P}/\partial t) \times \boldsymbol{B}$, the above Lorentz force predicts a total radiation force $\boldsymbol{F} = \boldsymbol{F}_e + \boldsymbol{F}_m = \tfrac{1}{4}\varepsilon_\mathrm{o}(n^2 - 1)\nabla(|E_x|^2 + |E_y|^2 + |E_z|^2)$. This force, which is directly proportional to the intensity gradient, has been used to explain the local compression of the liquid under a focused beam, and the subsequent formation of a bulge on the liquid surface.

The aforementioned expression for the electric component $\boldsymbol{F}_e$ of the Lorentz force, however, produces inconsistencies when applied to other problems, such as the problem of oblique incidence on a dielectric slab [5] or on a dielectric wedge [6]. We have shown in a series of recent articles [5-9] that the proper form of the Lorentz law that yields consistent results across the board is $\boldsymbol{F} = -(\nabla\cdot\boldsymbol{P})\boldsymbol{E} + (\partial\boldsymbol{P}/\partial t) \times \boldsymbol{B}$, where $-(\nabla\cdot\boldsymbol{P})$ is the density of bound charges within the dielectric medium. This form of the Lorentz law, when applied in conjunction with Maxwell's equations to the bulk as well as to the surface(s) of dielectrics, produces logically consistent results, but it also predicts novel behavior in certain cases (such as the case of a focused beam on a liquid surface) that requires verification through refined experiments.

In this paper we apply the latter form of the Lorentz law to the problem of radiation pressure when a polarized beam of light enters from the free-space into a transparent liquid (e.g., purified water). The schematic of the system under consideration is shown in Fig. 1; here a Gaussian beam of light having an elliptical cross-section enters a medium of refractive index $n$ at normal incidence. A fraction $\rho$ of the incident $E_x, E_y$ is reflected at the surface, while a fraction $\tau$ of the same field components is transmitted into the liquid. Maxwell's equations ensure the continuity of the electromagnetic field components $E_x$, $E_y$, $H_x$, $H_y$, $H_z$ at the interface. The discontinuity of $E_z$, however, gives rise to a (bound) surface charge density

$$\sigma_b(x, y, z = 0) = \varepsilon_0 \nabla \cdot \mathbf{E} = -\varepsilon_0 [1 - \rho - (\tau/n)] E_z^{(inc)}(x, y, z = 0), \qquad (1a)$$

which, when acted upon by the average $E_z$ at the interface,

$$<E_z(x, y, z = 0)> = \tfrac{1}{2}[1 - \rho + (\tau/n)] E_z^{(inc)}(x, y, z = 0), \qquad (1b)$$

produces a localized force $F_z^{(surface)}$. It is this force that tends to pull the liquid surface upward. (In the above equations, $E_z^{(inc)}(x, y, z = 0)$ is the incident $E_z$ in the $z = 0$ plane, which is also the waist of the Gaussian beam.) In Section 3 we derive an expression for $F_z^{(surface)}$ as a function of the various parameters of the incident beam.

Compared to the surface force, the electromagnetic forces within the body of the liquid are strong, but rather than contributing to the surface bulge, they appear to favor the formation of four vortices at the four corners of the beam. In Section 4 we derive a general expression for the volume force density inside the liquid, $\mathbf{F}(x, y, z > 0)$; the dependence of this force on the state of polarization of the incident beam will be examined in some detail.

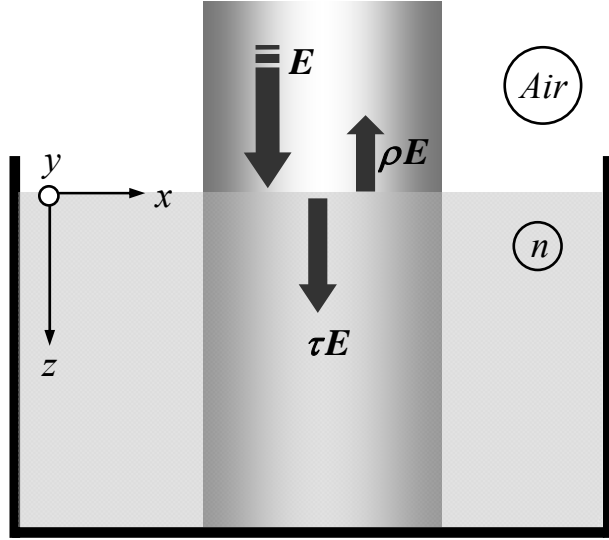

**Fig. 1**. A Gaussian beam having an elliptical cross-section (1/e radii $r_{xo}$, $r_{yo}$ along the $x$- and $y$-axes) arrives at normal incidence at the air–water interface (water's refractive index $= n$); the beam's waist is located at the water surface. The incident beam's state of polarization is defined by its $E$-field components $E_x$, $E_y$. In addition to the $x$- and $y$-components, both $E$- and $H$-fields have components along the $z$-axis, $E_z$ and $H_z$, given by Eqs. (3). Reflection and transmission coefficients (for $E_x$, $E_y$) are $\rho$ and $\tau$, respectively.

**2. Gaussian beam characteristics**. Consider a Gaussian beam of elliptical cross-section, propagating along the $z$-axis in an isotropic, homogeneous medium of refractive index $n$. In general, both $E$- and $H$-fields will have components along all three axes, namely,

$$\boldsymbol{E}(x,y,z) = \frac{\sqrt{\alpha(z)\beta(z)}}{\sqrt{\alpha(0)\beta(0)}} \exp[-\alpha(z)x^2 - \beta(z)y^2] \exp(i2\pi z/\lambda)(E_x\hat{\boldsymbol{x}} + E_y\hat{\boldsymbol{y}} + E_z\hat{\boldsymbol{z}}), \quad (2a)$$

$$\boldsymbol{H}(x,y,z) = \frac{\sqrt{\alpha(z)\beta(z)}}{\sqrt{\alpha(0)\beta(0)}} \exp[-\alpha(z)x^2 - \beta(z)y^2] \exp(i2\pi z/\lambda)[-(nE_y/Z_o)\hat{\boldsymbol{x}} + (nE_x/Z_o)\hat{\boldsymbol{y}} + H_z\hat{\boldsymbol{z}}]. \quad (2b)$$

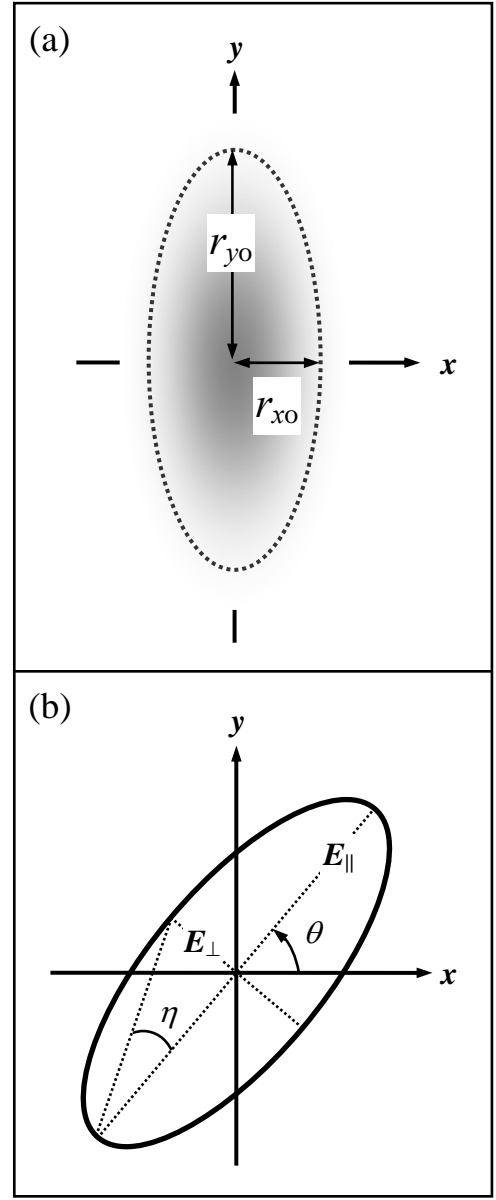

Here $E_x, E_y, E_z$ are the (complex) amplitudes of the $E$-field components at the origin $(x, y, z) = (0, 0, 0)$, $\lambda = \lambda_o/n$ is the wavelength inside the dielectric, $Z_o = \sqrt{\mu_o/\varepsilon_o} \approx 377\Omega$ is the impedance of the free-space, and $\alpha(z) = [1/r_x^2(z)] - i\pi/[\lambda R_x(z)]$ is the Gaussian beam's complex parameter defined in terms of the beam's $1/e$ radius $r_x(z)$ and its radius of curvature $R_x(z)$. The parameter $\alpha(z)$ evolves from its value of $\alpha(0)$ at the waist according to the formula $1/\alpha(z) = [1/\alpha(0)] + i\lambda z/\pi$. Similarly, $\beta(z) = [1/r_y^2(z)] - i\pi/[\lambda R_y(z)]$, with its corresponding evolution rule $1/\beta(z) = [1/\beta(0)] + i\lambda z/\pi$. At the waist the wavefront is flat, meaning that $R_x(0) = R_y(0) = \infty$; also for simplicity's sake we shall write the $1/e$ radii at the waist, $r_x(0), r_y(0)$, as $r_{xo}$ and $r_{yo}$; see Fig. 2(a).

The beam's $E$-field may be derived from the $H$-field using the Maxwell equation $\nabla \times \boldsymbol{H} = -i\omega\varepsilon_o\varepsilon\boldsymbol{E}$. Conversely, the beam's $H$-field may be derived from its $E$-field using $\nabla \times \boldsymbol{E} = i\omega\mu_o\boldsymbol{H}$. In particular, $E_z$ and $H_z$ may be determined in terms of $E_x, E_y$ as follows:

$$E_z = -i(\lambda/\pi)[\alpha(z)xE_x + \beta(z)yE_y], \quad (3a)$$

$$H_z = +i(\lambda/\pi)(n/Z_o)[\alpha(z)xE_y - \beta(z)yE_x]. \quad (3b)$$

**Fig. 2**. (a) Cross-sectional profile of the light amplitude distribution at the waist of a Gaussian beam described by Eqs. (2); the $1/e$ radii along the $x$- and $y$-axes are $r_{xo}$ and $r_{yo}$, respectively. (b) Ellipse of polarization at the beam's waist. The $E$-field components $E_x = |E_x|\exp(i\phi_x)$ and $E_y = |E_y|\exp(i\phi_y)$ along the coordinate axes determine the major and minor semi-axes of the ellipse, $E_\parallel$ and $E_\perp$, as well as its orientation angle $\theta$, through Eqs. (4). The ellipticity $\eta$ of the polarization state is given by $\eta = \arctan(E_\perp/E_\parallel)$.

At the beam's waist the state of polarization is generally elliptical, as shown in Fig. 2(b), with $E_x$ and $E_y$ related to the ellipse's semi-axes $E_\parallel, E_\perp$ and orientation angle $\theta$ as follows:

$$E_x = |E_x|\exp(i\phi_x) = E_\parallel \cos\theta + iE_\perp \sin\theta, \quad (4a)$$

$$E_y = |E_y|\exp(i\phi_y) = E_\parallel \sin\theta - iE_\perp \cos\theta. \quad (4b)$$

The ellipticity $\eta$ of the state of polarization is defined as $\eta = arctan(E_\perp/E_\parallel)$. The following useful relations are readily derived from Eqs. (4):

$$|E_x|^2 + |E_y|^2 = E_\parallel^2 + E_\perp^2, \tag{5a}$$

$$|E_x|^2 - |E_y|^2 = (E_\parallel^2 - E_\perp^2) \cos(2\theta), \tag{5b}$$

$$2 |E_x| |E_y| \cos(\phi_x - \phi_y) = (E_\parallel^2 - E_\perp^2) \sin(2\theta), \tag{5c}$$

$$\cos 2\eta = (E_\parallel^2 - E_\perp^2)/(E_\parallel^2 + E_\perp^2). \tag{5d}$$

The power $Q_o$ of the incident beam may be obtained by setting $n = 1.0$ in Eqs. (2), then integrating the $z$-component of the Poynting vector, $\boldsymbol{S} = \tfrac{1}{2}Real(\boldsymbol{E} \times \boldsymbol{H}^*)$, over the entire $xy$-plane at the waist (i.e., at $z = 0$). We find,

$$Q_o = \tfrac{1}{4} [\pi r_{xo} r_{yo}/Z_o] (E_\parallel^2 + E_\perp^2). \tag{6}$$

At the waist, the beam is reflected from the liquid surface, also located at $z = 0$; the reflection and transmission coefficients (for $E_x$, $E_y$) are given by

$$\rho = (1 - n)/(1 + n) \tag{7a}$$

$$\tau = 2/(1 + n) \tag{7b}$$

The above relations guarantee the continuity of $E_x$, $E_y$, $H_x$, $H_y$, $H_z$ across the boundary.

**3. Radiation force exerted on surface charges**. At the dielectric surface $E_z$ is discontinuous, although $D_z = \varepsilon_o \varepsilon E_z$ remains continuous. The induced (bound) surface charge density $\sigma_b$ is given by the discontinuity of $E_z$, as follows:

$$\sigma_b(x, y, z=0) = 2i \varepsilon_o (\lambda_o/\pi) [(n - n^{-1})/(n+1)] [\alpha(0) x E_x + \beta(0) y E_y] \exp[-\alpha(0) x^2 - \beta(0) y^2]. \tag{8}$$

The effective $E_z$ acting on these charges is the average of the $E_z$ just above and just below the interface, namely,

$$E_z(x, y, z=0) = -i (\lambda_o/\pi) [(n + n^{-1})/(n+1)] [\alpha(0) x E_x + \beta(0) y E_y] \exp[-\alpha(0) x^2 - \beta(0) y^2]. \tag{9}$$

In what follows $\alpha(0)$ and $\beta(0)$ will be replaced by their respective values, $(1/r_{xo})^2$ and $(1/r_{yo})^2$. The force density (i.e., force per unit area) at the air-liquid interface is thus given by

$$F_z(x,y,z=0) = \tfrac{1}{2}Real(\sigma_b E_z^*) = -\varepsilon_o (\lambda_o/\pi)^2 [(n^2 - n^{-2})/(n+1)^2]$$

$$\times [x^2|E_x|^2/r_{xo}^4 + y^2|E_y|^2/r_{yo}^4 + 2xy Real(E_x E_y^*)/(r_{xo}^2 r_{yo}^2)] \exp[-2(x/r_{xo})^2 - 2(y/r_{yo})^2]. \tag{10}$$

Integrating the force density of Eq. (10) over the $xy$-plane and using Eqs. (5) and (6), one finds

$$F_z^{(surface)} = -\tfrac{1}{4} (Q_o/c)[(n^2 - n^{-2})/(n+1)^2](\lambda_o/\pi)^2$$

$$\times \{(1/r_{xo})^2 + (1/r_{yo})^2 + [(1/r_{xo})^2 - (1/r_{yo})^2] \cos(2\theta) \cos(2\eta)\}, \tag{11}$$

where $c$ is the speed of light in vacuum. For example, when $Q_o = 1.0$W, $n = 1.335$, $\lambda_o = 0.512$μm, $r_{xo} = 1.5$μm, $r_{yo} = 2.8$μm, and $\eta = 0°$, we will have $F_z^{(surface)} = -(4.4\cos^2\theta + 1.26\sin^2\theta)$ pN. The surface force is thus strongest when the incident (linear) polarization vector is aligned with the narrow dimension of the beam ($\theta = 0°$), and declines continuously as the polarization vector rotates toward the long axis of the elliptical spot ($\theta = 90°$).

**4. Force distribution within the bulk**. Inside the liquid, the force density (i.e., force per unit volume) is given by the magnetic component of the Lorentz law,

$$\bm{F}(x, y, z > 0) = \tfrac{1}{2} Real(\bm{J}_b \times \bm{B}^*), \qquad (12)$$

where $\bm{B} = \mu_o \bm{H}$, and the bound current density $\bm{J}_b$ is the time-derivative of the polarization density $\bm{P}$, that is, $\bm{J}_b = -i\omega\varepsilon_o(\varepsilon - 1)\bm{E}$. Note that Eq. (12) does not include any contributions from the $E$-field component of the Lorentz force, namely, $\bm{F}_e = \rho_b \bm{E} = -(\nabla\cdot\bm{P})\bm{E}$, where $\rho_b$ is the bound charge density. The reason is that the first Maxwell equation, $\nabla\cdot\bm{D} = 0$, together with the constitutive relation $\bm{P} = (1 - \varepsilon^{-1})\bm{D}$ for isotropic, homogeneous dielectrics, ensure that $\nabla\cdot\bm{P} = 0$.

To determine the force density inside the liquid we substitute for $\bm{E}$ and $\bm{H}$ in Eq. (12) from Eqs. (2) and (3), and find, with the help of Eqs. (5)-(7),

$$\bm{F}(x, y, z > 0) = \frac{16\pi^3(n-1)(Q_o/c)\cos(2\eta)\exp\{-2[x/r_x(z)]^2 - 2[y/r_y(z)]^2\}}{(n+1)[(\pi r_{xo})^2 + (\lambda z/r_{xo})^2][(\pi r_{yo})^2 + (\lambda z/r_{yo})^2]}$$

$$\times \{[x\zeta^{-1}\cos(2\theta) + y\zeta\sin(2\theta)]\hat{\bm{x}} + [x\zeta^{-1}\sin(2\theta) - y\zeta\cos(2\theta)]\hat{\bm{y}}\}. \qquad (13)$$

Here $\zeta(z) = r_x(z)/r_y(z) = \sqrt{(\pi r_{xo})^2 + (\lambda z/r_{xo})^2}/\sqrt{(\pi r_{yo})^2 + (\lambda z/r_{yo})^2}$ is a dimensionless parameter. Note that the force density in the bulk of the liquid has no component along the $z$-axis. Also, for circularly polarized beams, $\eta = \pm 45°$ and the above force density reduces to zero, even though the surface force, given by Eq. (11), retains a finite value. In general, the force density of Eq. (13) is strongest at the waist ($z = 0$), and decreases with an increasing $z$ as the denominator increases. The explicit dependence of $\bm{F}$ on the refractive index $n$ of the liquid through the coefficient $(n-1)/(n+1)$ arises from the proportionality of $\bm{P}$ to $(n^2 - 1)$ as well as the dependence on $n$ of the power transmission coefficient $\tau^2$ at the liquid surface; see Eq. (7b).

In general, the only effect of polarization ellipticity $\eta$ on the force distribution of Eq. (13) is to reduce the strength of $\bm{F}$ by a factor of $\cos(2\eta)$. For maximum force, therefore, one must use linearly-polarized light. The direction of polarization $\theta$ enters Eq. (13) through the last term, which describes a vector field in the $xy$-plane; the length of the vector, $\sqrt{[x/\zeta(z)]^2 + [y\zeta(z)]^2}$, at any point $(x, y, z)$ within the liquid is independent of $\theta$. At any given depth $z$, both the magnitude and the direction of the force in the $xy$-plane vary with the $x, y$ coordinates. Rotating the incident polarization vector by $90°$, i.e., replacing $\theta$ with $\theta + 90°$ while keeping the other parameters of Eq. (13) constant, reverses the direction of $\bm{F}$ everywhere within the body of the liquid.

Figure 3 shows computed plots of $\bm{F}(x, y, z = z_o)$ in an $xy$-plane where $r_x(z_o) = 1.5$μm, $r_y(z_o) = 2.8$μm – corresponding to $\zeta(z_o) = 0.54$ – for several values of $\theta$. In each case the radiation pressure appears to favor the formation of four vortices in the four corners of the beam. While the strength of the driving force of these counter-rotating vortices does not depend on the orientation angle θ of the polarization vector, the sense of rotation at each point is very much a function of the polarization direction. For $Q_o = 1.0$ W, the magnitude of this force in water ($n = 1.335$) is of the order of $0.10$ nN/μm$^3$ in the vicinity of the beam's waist, which is substantially greater than the corresponding surface force.

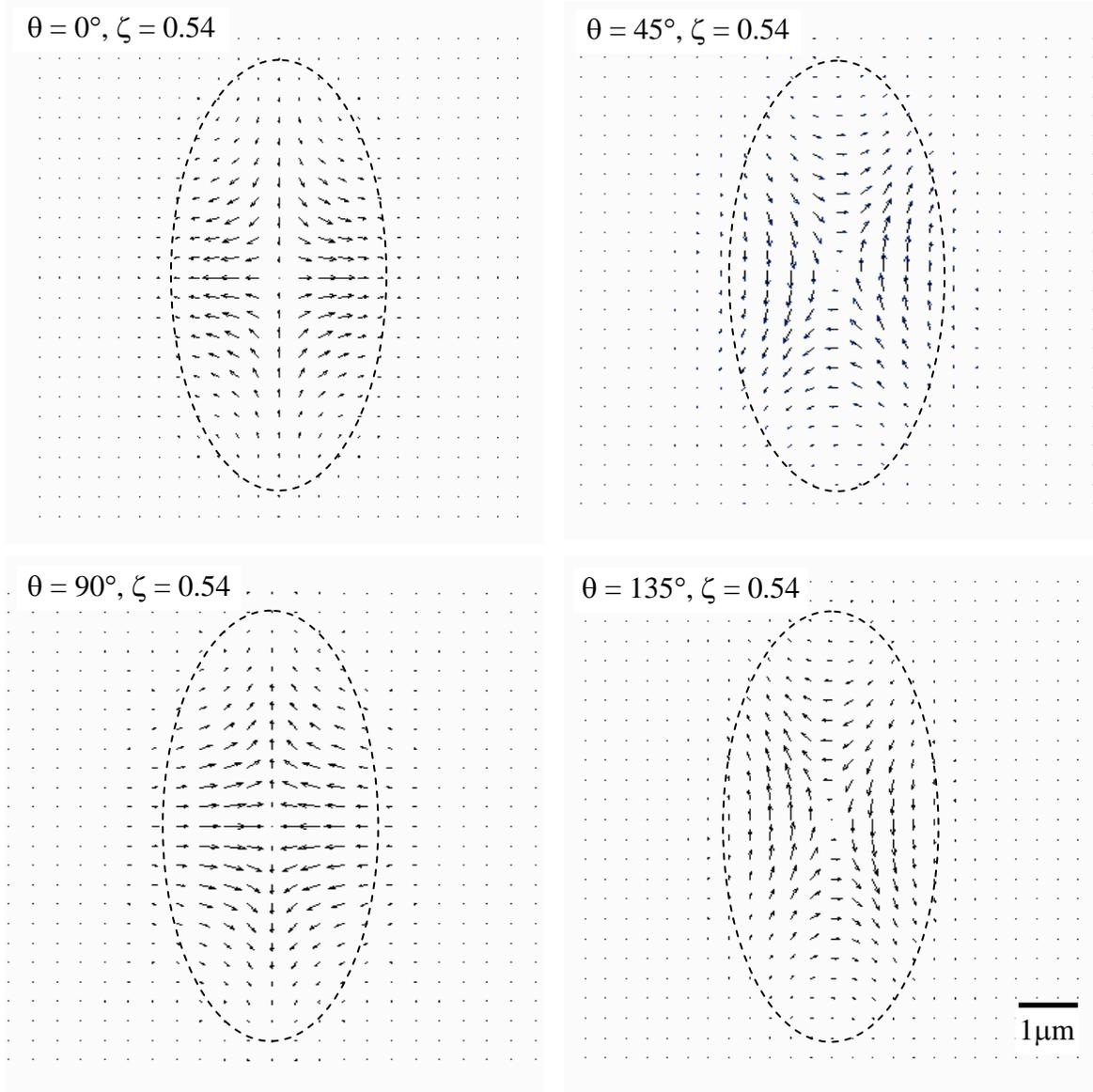

**Fig. 3**. Computed plots of the volume force density $F(x, y, z = z_o)$ in an *xy*-plane within the liquid for a Gaussian beam having $r_x(z_o) = 1.5$μm, $r_y(z_o) = 2.8$μm, for several values of the polarization orientation angle θ. In each case the radiation pressure appears to favor the formation of four vortices in the four corners of the beam. The polarization ellipticity $η$ does not affect the overall pattern of force distribution, but reduces the magnitude of $F$ everywhere by $\cos(2η)$.

**Acknowledgments.** The author is grateful to Pavel Polynkin, Ewan Wright, Brian Anderson, Armis Zakharian, and Rodney Loudon for many helpful discussions. This work is supported by the AFOSR contract F49620-02-1-0380 with the Joint Technology Office, by the *Office of Naval Research* MURI grant No. N00014-03-1-0793, and by the *National Science Foundation* STC Program under agreement DMR-0120967.

### References

1. A. Ashkin and J. Dziedzic, "Radiation pressure on a free liquid surface," Phys. Rev. Lett. **30**, 139-142 (1973).
2. R. Loudon, "Theory of the forces exerted by Laguerre-Gaussian light beams on dielectrics," Phys. Rev. A. **68**, 013806 (2003).
3. R. Loudon, "Radiation pressure and momentum in dielectrics," Fortschr. Phys. **52**, 1134-1140 (2004).
4. J. D. Jackson, *Classical Electrodynamics*, 2nd edition, Wiley, New York, 1975.
5. M. Mansuripur, "Radiation pressure and the linear momentum of the electromagnetic field," Opt. Express **12**, 5375-5401 (2004), http://www.opticsexpress.org/abstract.cfm?URI=OPEX-12-22-5375.
6. M. Mansuripur, A. R. Zakharian, and J. V. Moloney, "Radiation pressure on a dielectric wedge," Opt. Express **13,** 2064-2074 (2005), http://www.opticsexpress.org/abstract.cfm?URI=OPEX-13-6-2064.
7. M. Mansuripur, "Radiation pressure and the linear momentum of light in dispersive dielectric media," Opt. Express **13,** 2245-2250 (2005), http://www.opticsexpress.org/abstract.cfm?URI=OPEX-13-6-2245.
8. A. R. Zakharian, M. Mansuripur, and J. V. Moloney, "Radiation pressure and the distribution of electromagnetic force in dielectric media," Opt. Express **13,** 2321-2336 (2005), http://www.opticsexpress.org/abstract.cfm?URI=OPEX-13-7-2321.
9. M. Mansuripur, "Angular momentum of circularly polarized light in dielectric media," Opt. Express **13,** 5315-5324 (2005), http://www.opticsexpress.org/abstract.cfm?URI=OPEX-13-14-5315.